\documentclass{kapproc} 

\usepackage{procps}

\usepackage[dvips]{graphicx}

\upperandlowercase

\normallatexbib 

\begin{document}

\articletitle{The Josephson Bifurcation Amplifier for Quantum
Measurements}

\author{I. Siddiqi, R. Vijay, F. Pierre, C.M. Wilson, L. Frunzio, M. Metcalfe, C. Rigetti, and M.H. Devoret}
\affil{Departments of Applied Physics and Physics\\
Yale University, USA} \email{irfan.siddiqi@yale.edu}

\begin{abstract}
We have constructed a new type of amplifier whose primary purpose
is the readout of superconducting quantum bits. It is based on the
transition of an RF-driven Josephson junction between two distinct
oscillation states near a dynamical bifurcation point. The main
advantages of this new amplifier are speed, high-sensitivity, low
back-action, and the absence of on-chip dissipation. Using pulsed
microwave techniques, we demonstrate bifurcation amplification in
nanofabricated Al junctions and verify that the performance
predicted by theory is attained.
\end{abstract}

\section{Introduction}

Josephson first noted that the superconducting tunnel junction can
be viewed as a non-linear, non-dissipative, electrodynamic
oscillator \protect\cite{Josephson}. We exploit this non-linearity
to produce a new type of high-sensitivity amplifier, the Josephson
Bifurcation Amplifier (JBA). No shunt resistors are required in
our amplification scheme, and it is thus possible to take
advantage of the elastic character of the junction and eliminate
on-chip dissipation, thereby minimizing the back-action of the
amplifier. The combination of high-sensitivity and minimal
back-action makes the JBA well-suited for measurements on quantum
systems such as superconducting qubits, and make it a strong
candidate for reaching the quantum noise limit.

The operation of the JBA is represented schematically in Fig. 1.
The central element is a Josephson junction, shunted with a
lithographic capacitor, whose critical current $I_{0}$ is
modulated by an input signal (input port). Coupling between the
junction and the input signal can be achieved through different
schemes, examples of which involve placing the JBA in a SQUID loop
\cite{DELFT} or in parallel with a SSET \cite{Cottet}. The
junction is driven with a pure AC signal $i_{RF}\sin(\omega t)$ in
the microwave frequency range fed via a transmission line through
a circulator (drive port). In the underdamped regime, for certain
values of $\omega$ and $i_{RF}$, two possible oscillation states
which differ in amplitude and phase (denoted "0" and "1") can
coexist. The reflected component of the drive signal, measured
through another transmission line connected to the circulator
(output port), is a convenient signature of the junction
oscillation state. At the bifurcation point where switching
between oscillation states occurs, the system becomes infinitely
sensitive, in the absence of thermal and quantum fluctuations, to
variations in $I_{0}$. At finite temperature, the energy stored in
the oscillation can always be made larger than thermal
fluctuations by increasing the scale of $I_{0}$, thus preserving
sensitivity. Small variations in $I_{0}$ are transformed into
readily discernible changes in the escape rate $\Gamma_{01}$ from
state 0 to 1. Back-action is minimized in this arrangement since
the only fluctuations felt at the input port arise from the
fluctuations of the $50\,\Omega$ drive port whose dissipative
elements are physically separated from the junction via a
transmission line of arbitrary length and can therefore be
thermalized efficiently to base temperature. Additionally, the
frequency band over which the back-action contributes is narrow,
and well controlled.

In section 2, simplified expressions adapted from the theory of
activated escape in a driven non-linear oscillator \cite{Dykman}
are presented. Details of the devices and the measurement setup
are presented in Section 3. Experimental results are given in
Section 3, and concluding remarks are in Section 4.

\section{Theory}

The tunnelling of Cooper pairs manifests itself as a non-linear
inductance that shunts the linear junction self-capacitance
$C_{J}$, formed by the junction electrodes and the tunnel oxide
layer. The constitutive relation of the non-linear inductor can be
written as $I(t)=I_{0}\sin\delta\left(  t\right) $, where $I(t)$,
$\delta\left(  t\right)  =\int_{-\infty}^{t}dt^{\prime
}V(t^{\prime})/\varphi_{0}$ and $V(t)$ are the current,
gauge-invariant phase-difference and voltage corresponding to the
inductor, respectively, while the parameter $I_{0}$ is the
junction critical current. Here $\varphi_{0}=\hbar/2e$ is the
reduced flux quantum. The dynamics of the junction are given by
the time evolution of $\delta$, which exhibits the
motion of a phase particle in a cosine potential $U(\delta)=-\varphi_{0}%
I_{0}\cos(\delta)$. For small oscillation amplitude about the
potential minima, the frequency of oscillation is given for zero
DC bias current by the plasma frequency
$\omega_{P0}=1/\sqrt{L_{J}C_{J}}$ where $L_{J}=\varphi _{0}/I_{0}$
is the effective junction inductance. As the oscillation amplitude
increases, the potential "softens" and $\omega_{P}$ decreases, an
effect which has been measured in both the classical and quantum
regime \cite{Devoret,Martinis,Yurke1,Holst}. A more dramatic
non-linear effect manifests itself if the junction is driven with
an AC current $i_{RF}\sin\omega t$ at a frequency $\omega$
slightly below $\omega_{P0}$. If the quality factor
$Q=C_{J}\omega_{P0}/Re[Z^{-1}(\omega_{P0})]$ is greater than
$\sqrt{3}/2\alpha$, where $Z(\omega_{P0})$ is the impedance of the
junction electrodynamic environment and
$\alpha=1-\omega/\omega_{P0}$ the detuning parameter, then the
junction switches from one dynamical oscillation state to another
when $i_{RF}$ is ramped above a critical value $I_{B}$
\cite{Landau}. For $i_{RF}<I_{B}$, the oscillation state is
low-amplitude and phase-lagging while for $i_{RF}>I_{B}$, the
oscillation state is high-amplitude and phase-leading. This
generic non-linear phenomenon, which we refer to as "dynamical
switching", is reminiscent of the usual "static switching" of the
junction from the zero-voltage state to the voltage state when the
DC current bias exceeds the critical current $I_{0}$
\cite{Fulton}. However, an important distinction between dynamical
and static switching is that in dynamical switching, the phase
particle remains confined to only one well of the junction cosine
potential, and the time-average value of $\delta$ is always zero.
The junction never switches to the voltage\thinspace\ state, and
thus no DC voltage is generated. Also, for dynamical switching,
the current $I_{B}$ depends both on $Q$ and on the detuning
$\alpha$.

\begin{figure}[b]
\includegraphics[width=4.8in]{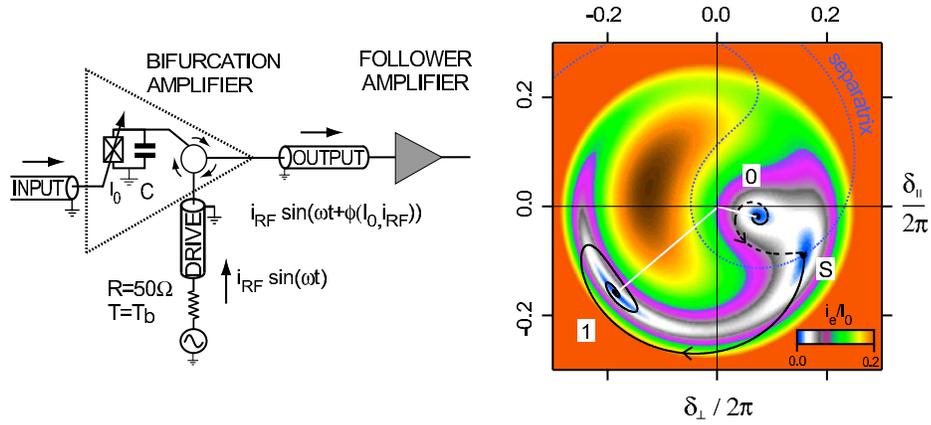}
\sidebyside {\caption{Schematic diagram of the Josephson
bifurcation amplifier. A junction with critical current $I_{0}$,
parametrically coupled to the input port, is driven by an RF\
signal which provides the power for amplification. In the vicinity
of the dynamical bifurcation point $i_{RF}=I_{B}$, the reflected
signal phase $\phi$ depends critically on the input signal.}}
{\caption{Poincare section of an RF-driven Josephson junction in
the bistable regime $(\alpha=(1-\omega
/\omega_{p})=0.122,i_{RF}/I_{B}=0.87)$. The two stable oscillation
states, labelled by 0 and 1, are indicated by white line segments.
Point S which lies on the separatrix is the saddle point at which
the escape trajectory from state 0 (dashed line) meets the
retrapping trajectory into state 1 (solid line). }}
\end{figure}

In presence of the microwave drive $i_{RF}\sin(\omega t)$, the
oscillations in the junction phase can be parameterized using
in-phase and quadrature phase
components $\delta(t)=\delta_{\parallel}\sin(\omega t)+\delta_{\perp}%
\cos(\omega t)$ (higher harmonics of oscillation are
negligible).The two oscillation states appear as two points in the
$\left(  \delta_{\parallel },\delta_{\perp}\right)  $ plane and
are denoted by vectors labelled 0 and 1 (see Fig. 2). The
error-current \cite{Kautz} which describes the generalized
force felt by the system is also plotted in the $\left(  \delta_{\parallel}%
,\delta_{\perp}\right)  $ plane. Its value goes to zero at the
attractors corresponding to states 0 and 1 and also at a third
extremum which is the dynamical saddle point. Also shown in Fig. 2
is the calculated escape trajectory \cite{Dykmanescape} from state
0 (dashed) and the corresponding retrappping trajectory
\cite{Dmitriev} into state 1 (solid line). Fig. 2 has been
computed for $\alpha=0.122$, $Q=20$ and $i_{RF}/I_{B}=0.87$. These
values correspond to typical operating conditions in our
experiment. The dynamical switching from state 0 to 1 is
characterized by a phase shift given here by
$\mathrm{tan}^{-1}\left[  \left(
\delta_{\parallel}^{1}-\delta_{\parallel
}^{0})/(\delta_{\perp}^{1}-\delta_{\perp}^{0}\right)  \right]
=-139\,\mathrm{deg}$. Using the junction phase-voltage
relationship and the transmission line equations, we can calculate
the steady-state magnitude and phase of the reflected microwave
drive signal. The change in the oscillation of $\delta$ results in
a shift of the reflected signal phase $\Delta\phi
_{01}=89\,\mathrm{deg}$. Since there is no source of dissipation
in the junction chip, there should be no change in the magnitude
of the reflected signal power, even though
$\sqrt{(\delta_{\parallel}^{1}-\delta_{\parallel
}^{0})^{2}+(\delta_{\perp}^{1}-\delta_{\perp}^{0})^{2}}\neq0$.

Both static and dynamical switching can be described by an
Arhennius law in which the escape rate
$\Gamma_{01}=(\omega_{att}/2\pi)\exp(-\Delta U/k_{b}T)$ is written
as the product of an attempt frequency $\omega_{att}/2\pi$ and a
Boltzman factor which contains the potential barrier height
$\Delta U$ and the system temperature $T$. For the case of a DC
current bias, the cosine potential, near the switching
point, is approximated as a cubic potential with height $\Delta U_{st}=(4\sqrt{2}/3)\,\,\varphi_{0}%
I_{0}\,\left(  1-i_{DC}/I_{0}\right)  ^{3/2}$ where $i_{DC}$ is
the bias current. The attempt frequency is the plasma frequency
$\omega_{P}.$ In the absence of fluctuations, the characteristic
current at which switching occurs is $I_{0}.$ For the AC driven
junction, the dynamical switching from oscillation state 0 to 1
can be cast in a similar form using the model of a particle in a
cubic metapotential \cite{Dykman}. In this case, the effective
barrier height is, to lowest order in $1/(\alpha Q)$, $\Delta U_{dyn}%
=\,u_{dyn}(1-(i_{RF}/I_{B})^{2})^{3/2}$ with $u_{dyn}%
=64\hbar/(18e\sqrt{3})\,\,I_{0}\,\,\alpha(1-\alpha)^{3}$. The
attempt
frequency in the metapotential is given by $\omega_{a}=\omega_{a0}%
\,(1-(i_{RF}/I_{B})^{2})^{1/2}$ with
$\omega_{a0}=4/(3\sqrt{3}\,RC)\,(\omega _{p}-\omega)^{2}$. The
bifucrcation current $I_{B}$ where the 0 state ceases to exist is
given by $I_{B}=16/(3\sqrt{3})$
$\alpha^{3/2}(1-\alpha)^{3/2}\,I_{0}$.

\section{Devices and Setup}

Typical junction fabrication parameters limit the plasma frequency
to the 20 - 100 GHz range where techniques for addressing junction
dynamics are inconvenient. We have chosen to shunt the junction by
a capacitive admittance to lower the plasma frequency by more than
an order of magnitude and attain a frequency in 1-2 GHz range
(microwave L-band). \ In this frequency range, a simple on-chip
electrodynamic environment with minimum parasitic elements can be
implemented, and the hardware for precise signal generation and
processing is readily available. In the first step of sample
fabrication, a metallic underlayer -- either a normal metal (Au,
Cu) or a superconductor (Nb) -- was deposited on a silicon
substrate to form one plate of the shunting capacitor, followed by
the deposition of an insulating $\mathrm{Si}_{3}\mathrm{N}_{4}$
layer. Using e-beam lithography and double-angle shadow mask
evaporation, we subsequently fabricated the top capacitor plates
along with a micron sized
$\mathrm{Al}/\mathrm{Al}_{2}\mathrm{O}_{3}/\mathrm{Al}$ tunnel
junction. The critical current of the junction\ was in the range
$I_{0}=1-2\,\mathrm{\mu A}$. By varying both the dielectric layer
thickness and the pad area, the capacitance $C$ was varied between
$16$ and $40\,\mathrm{pF}$.

The junction + capacitor chip is placed on a microwave
circuit-board and is wire-bonded to the end of a coplanar
stripline which is soldered to a coaxial launcher affixed to the
side wall of the copper sample box. We anchor the RF leak-tight
sample box to the cold stage of a $^{3}\mathrm{He}$ refrigerator
with base temperature $T=280\,\mathrm{mK}$. The measurement setup
is schematically shown in Fig. 3. Microwave excitation signals are
coupled to the sample via the -13 dB side port of a directional
coupler. The reflected microwave signal passes through the direct
port of the coupler, and is amplified first using a cryogenic
$1.20-1.85\,\mathrm{GHz}$ HEMT amplifier with noise temperature
$T_{N}=\mathrm{4\,K}$. A DC bias current can be applied to the
junction by way of a bias tee. We use cryogenic attenuators,
isolators, and specially developed dissipative microstrip filters
on the microwave lines in addition to copper-powder and other
passive filters \cite{Martinis} on the DC lines to shield the
junction from spurious electromagnetic noise. In the first set of
experiments which probe the plasma resonance, a vector network
analyzer was used to both source a CW microwave signal and to
analyze the reflected power \cite{Siddiqi}. The dynamics of the
transition between the two oscillation states was then probed
using microwave pulses \cite{SiddiqiII}, generated by the
amplitude modulation of a CW source with a phase-locked arbitrary
waveform generator with $1\,\mathrm{ns}$ resolution. For the
pulsed experiments, the reflected signal was mixed down to
$100\,\mathrm{MHz}$ and digitally demodulated using a
$2\,\mathrm{GS/s}$ digitizer to extract the signal phase $\phi$.

\begin{figure}[ptbh]
\center\includegraphics[width=3.4in]{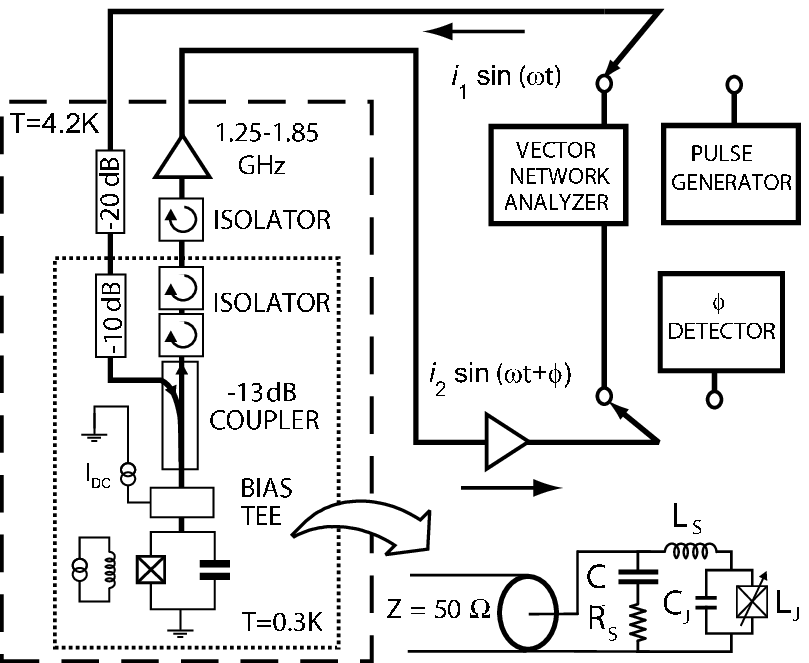}\caption{Schematic of
the measurement setup. Thick lines correspond to $50\,\Omega$
coaxial transmission lines. The network analyzer is used for CW
measurements. For probing the dynamics, the sample is switched to
the pulse generator and phase detector. A lumped element model for
the junction chip and measurement line is shown in the lower
right. The two ideal current sources actually represent external sources.}%
\label{FigSampleWL}%
\end{figure}

\section{Results}

We first probed the drive current dependence of the reflected
signal phase $\phi\left(  i_{RF}\right)  $ by applying a
$4\,\mathrm{\mu s}$ long symmetric triangular shaped pulse with a
peak value $0.185\,I_{0}$. The demodulated reflected signal was
divided into $20\,\mathrm{ns}$ sections, each yielding one
measurement of $\phi$ for a corresponding value of $i_{RF}$. The
measurement was repeated $6\times10^{5}$ times to obtain a
distribution of $\phi(i_{RF})$. In Fig. 4, the mode of the
distribution is plotted as a function of $i_{RF}/I_{0}$. For
$i_{RF}/I_{0}<0.125$, the bifurcation amplifier is always in state
0, $\phi$ is constant and assigned a value of
$0\,\mathrm{deg}$. As the drive current is increased above $i_{RF}%
/I_{0}=0.125$, thermal fluctuations are sufficiently large to
cause transitions to the 1 state. In the region between the two
dashed lines at $i_{RF}/I_{0}=0.125$ and $i_{RF}/I_{0}=0.160$,
$\phi$ displays a bimodal distribution with peaks centered at $0$
and $74\,\mathrm{deg}$ with the latter corresponding to the
amplifier in the 1 state. The dotted line in Fig. 4 is the average
reflected signal phase $\left\langle \phi\right\rangle $. When
$i_{RF}/I_{0}$ is increased above $0.160$, the system is only
found in state 1. In the decreasing part of the $i_{RF}$ ramp, the
system does not start to switch back to state 0 until
$i_{RF}/I_{0}=0.065$. The critical switching currents $I_{B}$ for
the $0\rightarrow1$ transition and $I_{\bar{B}}$ for the
$1\rightarrow0$ transition, calculated from numerical simulations
to treat the inductance of wire bonds, are denoted with ticks in
Fig. 4, and are in good agreement with experiment. The hysteresis
$I_{\bar{B}}<I_{B}$ is a consequence of the asymmetry in the
escape barrier height for the two states. Thus, the
$0\rightarrow1$ transition at $i_{RF}=I_{B}$ is nearly
irreversible, allowing the bifurcation amplifier to latch and
store its output during the integration time set by the
sensitivity of the follower amplifier.

To determine the sensitivity of the bifurcation amplifier, we have
characterized in detail the switching in the vicinity of the
$0\rightarrow1$ transition. We excited the system with two
different readout pulse protocols. In the first protocol, the
drive current was ramped from 0 to its maximum value in
$40\,\mathrm{ns}$ and was then held constant for $40\,\mathrm{ns}$
before returning to 0. Only the final $20\,\mathrm{ns}$ of the
constant drive period were used to determine the oscillation phase
with the first $20\,\mathrm{ns}$ allotted for settling of the
phase. Histograms taken with a $10\,\mathrm{MHz}$ acquisition rate
are shown in Fig. 5. In the upper panel, the two peaks
corresponding to states 0 and 1 can easily be resolved with a
small relative overlap of $10^{-2}$. The width of each peak is
consistent with the noise temperature of our HEMT amplifier. In
this first method, the latching property of the system has not
been exploited. In our second protocol for the readout pulse, we
again ramp for $40\,\mathrm{ns}$ and allow a settling time of
$20\,\mathrm{ns}$, but we then reduce the drive current by $20\%$
and measure the reflected signal for $300\,\mathrm{ns}$. In that
latter period, whatever state was reached at the end of the
initial $60\,\mathrm{ns}$ period is "latched" and time is spent
just increasing the signal/noise ratio of the reflected phase
measurement. As shown in the lower panel of Fig. 5, the two peaks
are now fully separated, with a relative overlap of
$6\times10^{-5}$ allowing a determination of the state 1
probability with an accuracy better than $10^{-3}$. This second
protocol would be preferred only for very precise time-resolved
measurements of $I_{0}$ or for applications where a low-noise
follower amplifier is impractical.

\begin{figure}[b]
\includegraphics[width=4.8in]{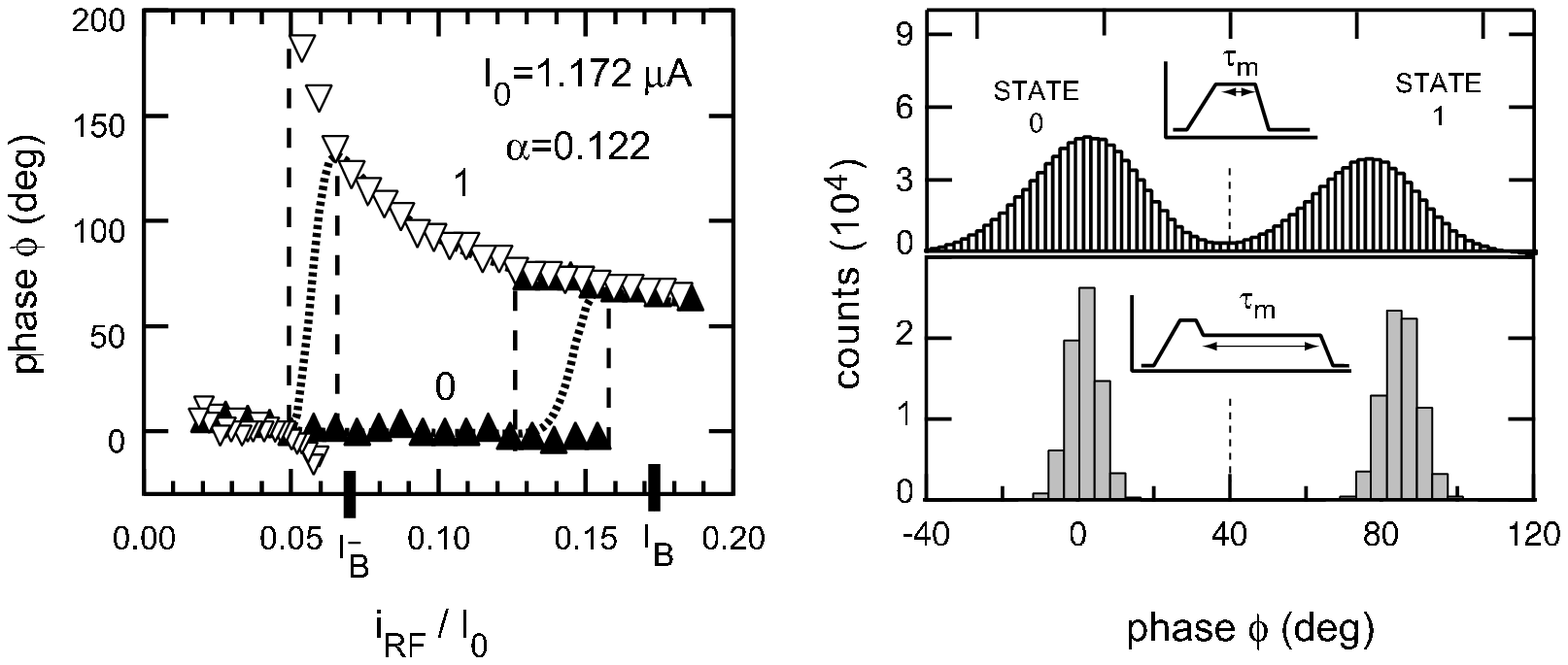}
\sidebyside {\caption{Hysteretic variation of the reflected signal
phase $\phi$ with drive current $i_{RF}/I_{0}$. Symbols denote the
mode of $\phi$, with up and down triangles corresponding to
increasing and decreasing $i_{RF}=I_{B}$, respectively. The dotted
line is $\left\langle \phi\right\rangle $. The calculated
bifurcation points, $I_{\bar{B}}$ and $I_{B}$, are marked on the
horizontal axis. The 0 and 1 phase states are reminiscent of the
superconducting and dissipative states of the DC current biased
junction.}} {\caption{Histograms of the reflected signal phase
$\phi$ at $i_{RF}/I_{0}=0.145$. The upper histogram contains
$1.6\times10^{6}$ counts with a measurement time
$\tau_{m}=20\,\mathrm{ns}$. The lower panel, taken with the
latching technique, has $1.5\times10^{5}$ counts with a
measurement time $\tau_{m}=300\,\mathrm{ns}$. Data here has been
taken under the same operating conditions as in Fig 4. The dashed
line represents the discrimination threshold between the 0 and 1
state.}}
\label{FigSampleWL}%
\end{figure}

A third experiment was performed to study the state 1 switching
probability $P_{01}\left(  i_{RF}\right)  $ for different values
of the temperature $T$ and $I_{0}$, the latter being varied with a
magnetic field applied parallel to the junction plane. Using the
first readout protocol and the discrimination threshold shown in
Fig. 5, we obtain the switching probability curves shown in Fig.
6. Defining the discrimination power $d$ as the maximum difference
between two switching probability curves which differ in $I_{0}$
we find that at $T=280\,\mathrm{mK}$, $d=57\%$ for $\Delta
I_{0}/I_{0}=1\%$ -- the typical variation observed in a
superconducting charge-phase qubit \cite{Vion}. The switching
probability curves should shift according to $(\Delta
I_{B}/I_{B})/(\Delta I_{0}/I_{0})=3/(4\alpha )-1/2+O(1/(\alpha
Q)^{2})$, which for our case takes the value 5.6. In Fig. 6, the
curves are shifted by $6\%$, which agrees well with this
prediction. For the case of the DC current biased junction,
similar curves would shift only by $1\%$ since the switching
current is $I_{0}$ itself. Comparable discrimination power using
DC switching has only been achieved in these devices at $T\leq
60\,\mathrm{mK.}$ As the temperature is increased, the switching
probability curves broaden due to increased thermal fluctuations
and the discriminating power decreases: at $T=480\,\mathrm{mK}$,
$d=49\%$.

Finally, we determined the escape rate
$\Gamma_{01}(i_{RF},I_{0},T)$ as a function of $i_{RF}$ by
measuring the time dependence of the switching probability, using
a method previously applied to the determination of the static
switching rates to the voltage state \cite{Turlot}. After the
initial ramp ($40\,\mathrm{ns}$) and settling period
($20\,\mathrm{ns}$), the reflected signal phase was extracted
every $20\,\mathrm{ns}$ for a duration of $1\,\mathrm{\mu s}$. By
repeating this measurement, we generated switching probability
histograms which we analyzed as $P_{01}\left(  t\right)
=1-\exp(-\Gamma_{01}\cdot t)$. To obtain the escape rate at
different temperatures, two different techniques were used. In the
first method, we varied the temperature of the cryostat and used a
magnetic field to keep the critical current constant at
$I_{0}=1.12\,\mathrm{\mu A}$. In the second method, $I_{0}$ was
kept fixed at $1.17\,\mathrm{\mu A}$, and a $1-2\,\mathrm{GHz}$
white noise source irradiating the junction was used to increase
the effective temperature. In Fig. 7 we show the drive current
dependence of the escape rate as
$(\ln(2\pi\omega_{a}/\Gamma_{01}))^{2/3}$ plotted versus
$i_{RF}^{2}$ for two different sample temperatures. Data in this
format should fall on a straight line with a slope $s\left(
T\right)  $ proportional to $\left(  u_{dyn}/k_{B}T\right)
^{2/3}$. A trace taken at $T=500\,\mathrm{mK}$ is also shown in
Fig 7.

    In parallel with these dynamical switching measurements, we
ran static switching measurements to  obtain an escape temperature
$T_{st}^{esc}$. Due to insufficient filtering in our RF amplifier
line outside the measurement band, $T_{st}^{esc}$
exceeded $T$ by $60\,\mathrm{mK}$. Using $u_{dyn}%
^{calc}$ and $s(T)$ we can cast the results of the dynamical switching
measurements into a dynamical escape temperature $T_{dyn}^{esc}=u_{dyn}^{calc}%
/k_{B}\,s\left(  T\right)  ^{3/2}$. We plot $T_{dyn}^{esc}$ versus
$T_{st}^{esc}$ in the inset of Fig. 7. The agreement is very good,
and only deviations at the highest temperatures are observed.$\ $
Analyzing the dynamical switching data with $T_{st}^{esc}$ in
place of $T$, we extract a value of $u_{dyn}=10.7\,\mathrm{K}$
from the $T=280\,\mathrm{mK}$ data with $I_{0}=1.17\,\mathrm{\mu
A}$ while the calculated value keeping higher order terms in
$1/\alpha Q$ is $u_{dyn}^{calc}=11.0\,\mathrm{K}$.

\begin{figure}[t]
\includegraphics[width=4.8in]{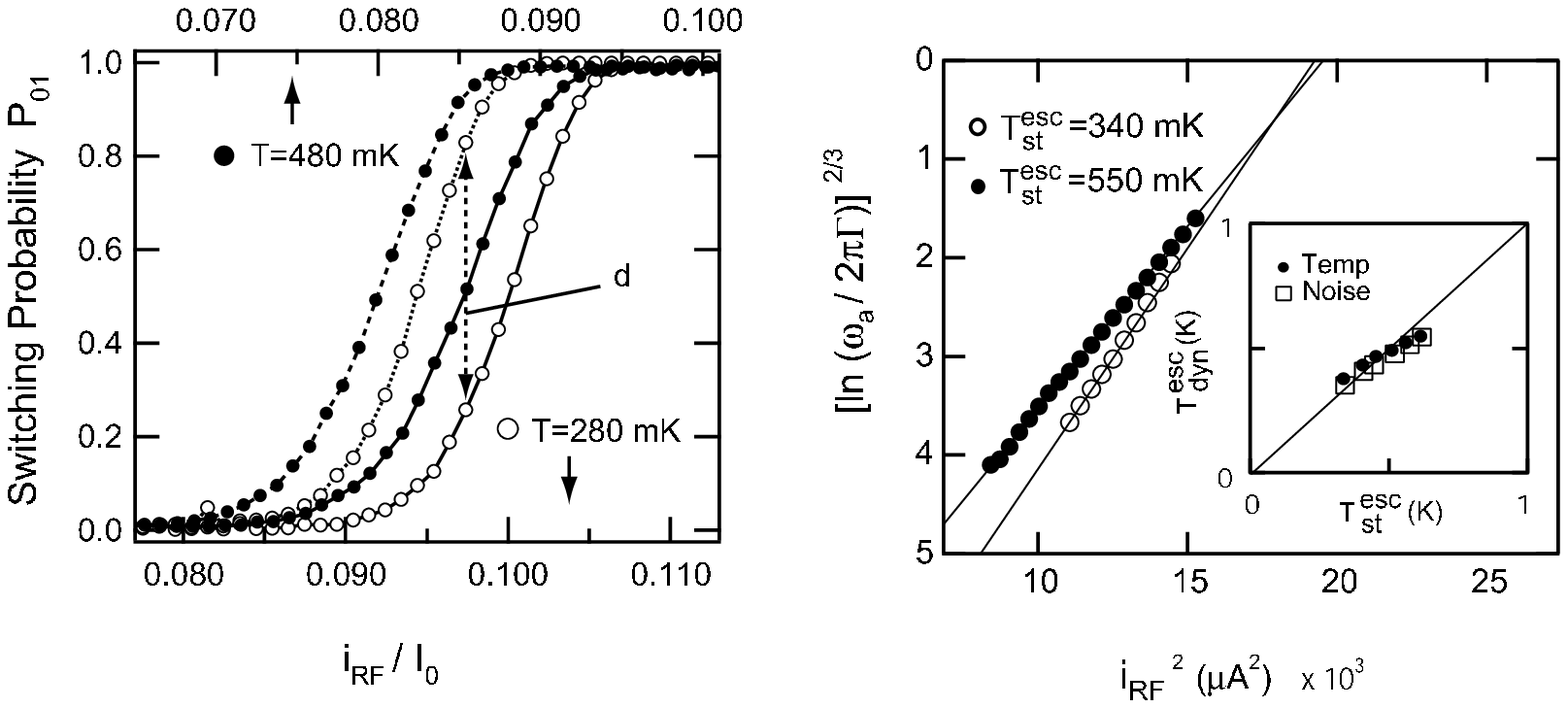}
\sidebyside {\caption{Switching probability curves at
$T=280\,\mathrm{mK}$ and $T=480\,\mathrm{mK}$ as a function of the
drive current $i_{RF}$. The discrimination power d is the maximum
difference between two curves at the same temperature which differ
by approximately $1\%$ in $I_{0}$.}} {\caption{Escape rate as a
function of
drive power for two different operating temperatures with $I_{0}%
=1.12\,\mathrm{\mu A}$. The inset shows the relationship between
dynamic and static escape temperatures when varying either the
sample temperature or the injected noise power.}}
\end{figure}

\section{Conclusion}

With the JBA operating at $T_{st}^{esc}=340\,\mathrm{mK}$, it is
possible to resolve with a signal/noise ratio of 1 a
$10\,\mathrm{nA}$ variation in $I_{0}$ in a total time $\leq80$
$\,\mathrm{ns}$, corresponding to a critical current sensitivity
of $S_{I_{0}}^{1/2}=3.3\times10^{-6}\,\mathrm{A\cdot Hz}^{-1/2}$.
This value is in agreement with the prediction from the
analytical theory $S_{I_{0}}^{1/2}=\eta(i_{RF}/I_{0},\alpha)\,(k_{B}%
T/\varphi_{0})\cdot\tau_{m}^{1/2}$, where $\eta\approx1.4$ near
the bifurcation point and $\varphi_{0}=\hbar/2e$. The advantage of
the bifurcation amplifier over SQUIDs \cite{Clarke} resides in its
extremely low back-action. Since there is no on-chip dissipation,
the only source of back-action is the matched isolator load, which
is efficiently thermalized at $T=280\,\mathrm{mK}$. An important
point is that in the JBA, only fluctuations from the load that
occur in a narrow band centered about the plasma frequency
contribute to the back-action, whereas in the SQUID noise from
many high frequency bands is also significant. Finally, the
bifurcation amplifier does not suffer from quasiparticle
generation associated with hysteretic SQUIDS \cite{DELFT} and DC\
current-biased junctions \cite{Cottet} which switch into the
voltage state. Long quasiparticle recombination times at low
temperatures limit the acquisition rate of these devices while the
recombination process itself produces excess noise for adjacent
circuitry \cite{Lukens}.

In conclusion, the JBA is competitive with the SQUID for
applications where low back-action is required. Its speed,
suppression of on-chip dissipation, and latching make
it ideal for the readout of superconducting qubits. At temperatures such that $T_{dyn}^{esc}%
\leq60\,\mathrm{mK}$, the discrimination power would be greater
than 95\%, hence permitting stringent tests of Quantum Mechanics,
like the violation of Bell's inequalities.

\begin{acknowledgments}
We would like to thank D. Prober, E. Boaknin, M. Dykman, L.
Grober, D. Esteve, D. Vion, S. Girvin and R. Schoelkopf for
discussions and assistance. This work was supported by the ARDA
(ARO Grant DAAD19-02-1-0044),the NSF (Grant ITR DMR-0325580,
DMR-0072022), and the Keck foundation.
\end{acknowledgments}

\end{document}